\documentclass[prl,showpacs,twocolumn]{revtex4}
\usepackage{graphicx}
\begin{document}

\title{Universal Scaling of $p_T$ Distribution of Particles
in Relativistic Nuclear Collisions}
\author{L.L. Zhu and C.B. Yang}
\affiliation{Institute of Particle Physics, Hua-Zhong Normal University, Wuhan 430079,
P.R. China}

\begin{abstract}
With the experimental data from STAR, PHENIX and BRAHMS on the centrality and rapidity
dependence of the $p_T$ spectrum in Au+Au and d+Au collisions, we show that
there exists a scaling distribution which is independent of the colliding system,
centrality and rapidity. The parameter for the average transverse momentum
$\langle p_T\rangle$ increases from peripheral to central d+Au collisions.
This increase accounts for the enhancement of particle production in those collisions.
A non-extensive entropy is used to derive the scaling function.

\pacs{25.75.Dw}
\end{abstract}

\maketitle

The distributions of produced particles play very important role in the study of
new state of matter produced in relativistic nuclear collisions. From the relationship
between the distributions at different centralities and rapidities one can see whether
there is a suppression or enhancement, thus dynamical mechanism for those particle
production can be revealed. In earlier works \cite{hy1}, a scaling distribution
for produced pions in Au+Au collisions in mid-rapidity region at RHIC energies
was found that is independent of the centrality and colliding energy. Similar
scaling behaviors have been found in \cite{zscal,cgc}. The existence of such scaling
behaviors of the spectrum for a species of particle may be an indication of universal
underlying particle production mechanism in the collisions.
One may ask two questions now. (1) Can the scaling be extended to non-central rapidity
region? (2) Can a scaling behavior similar to that shown in \cite{hy1} be
found for d+Au collisions at different centralities and rapidities? If the
answers to above two questions are yes, one should investigate the difference between
the scaling functions for different colliding systems.
In this paper, we look for the scaling behavior of particle
distributions in Au+Au collisions in noncentral rapidity region, in d+Au collisions
at different centralities and rapidities at RHIC energies. The same universal
scaling behavior is found for both Au+Au and d+Au collisions at different
centralities and rapidities. A possible origin for the universal
scaling distribution is given from the principle of maximum non-extensive entropy.
\begin{figure}[tbph]
\includegraphics[width=0.4\textwidth]{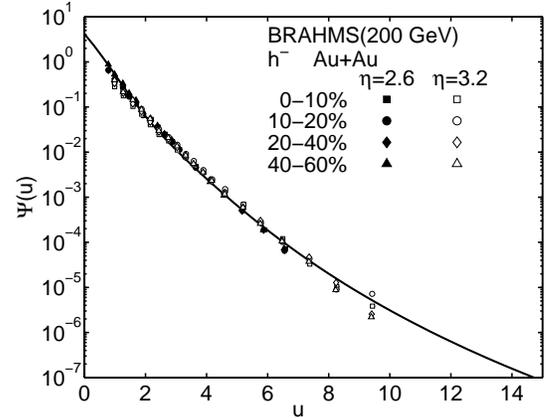}
\caption{Scaling distribution $\Psi(u)$ showing the coalescence of four rapidity
bins in the BRAHMS data for $h^-$ production at $\eta=2.6$ and 3.2 in Au+Au collisions.
The data points are taken from Fig.1 in \cite{exp1}.}
\label{fig1}
\end{figure}

Generally, the invariant particle distribution $E\frac{d^3N}{dp^3}$ for a given
species of particles depends on the collision centrality $\beta$, (pseudo)rapidity $y$
(or $\eta$) and the transverse momentum $p_T$. When we focus on the particle production
in a collision with given centrality in a small rapidity region around $y$, we should
distinguish two different factors. One is the number density $dN/dy$ in rapidity,
which tells us how many particles are produced in the region. The other is the probability
for a produced particle having transverse momentum $p_T$. It is this probability
that dictates the shape of the transverse momentum distributions. In this paper, we are
interested only in such a probability which can be defined as
\begin{equation}
\Phi(p_T,y,\beta)\equiv 2\pi\left.E\frac{d^3N}{dp^3}\right/\frac{dN}{dy}\ ,
\label{eq1}
\end{equation}
where $\frac{dN}{dy}=2\pi\int E\frac{d^3N}{dp^3} p_Tdp_T$ is the multiplicity of particles
produced in a unit rapidity region. So $\Phi(p_T,y,\beta)$
satisfies the normalization condition
\begin{equation}
\int \Phi(p_T,y,\beta) p_Tdp_T=1\ .
\label{norm}
\end{equation}
In the definition of $\Phi$, the centrality and rapidity dependence is written explicitly.
Experimental data show similar shape of the distributions for different centralities
and rapidities. Because of the similarity in the shape, those distributions of pions
produced in Au+Au collisions in the mid-rapidity region at difference centralities have
been put together in \cite{hy1} to the same curve by shrinking $p_T$ by a factor $K$ and
shifting the distribution by another factor $A$. Then a scaling variable is defined as
$u=p_T/\langle p_T\rangle$, and the normalized distribution is obtained as
a function $\Psi(u)=\langle p_T\rangle^2\Phi(\langle p_T\rangle u,y,\beta)$.
Here the average transverse momentum is defined as
\begin{equation}
\langle p_T\rangle =\int \Phi(p_T,y,\beta) p^2_Tdp_T\ .
\label{ave}
\end{equation}
The scaling function reads
\begin{eqnarray}
\Psi(u)=2.1\times10^4(u^2+11.65)^{-4.8}
[1+25\exp(-1.864u)] .
\label{scal}
\end{eqnarray}
The obtained $\Psi(u)$ is universal in the sense that it is independent of the
centrality $\beta$ at mid-rapidity in Au+Au collisions.

Now we extend the discussion to the particle production in non-central rapidity region
in Au+Au collisions at 200 GeV. For this purpose, we work on the experimental data
\cite{exp1} for negatively charged particles $h^-$ at large $\eta$ with exactly the same
procedures as in \cite{hy1} and try to put all the data points to the curve for the
scaling function $\Psi(u)$. It turns out that the scaling function given by
Eq. (\ref{scal}) can also describe the charged particle production at $\eta=$2.6 and
3.2 for different centralities, as can be seen in Fig. \ref{fig1}. The shrinking factor
for $p_T$ is given in TABLE \ref{tab1}.
\tabcolsep 0.05in
\renewcommand\arraystretch{1.1}
\begin{table}[tbph]
\begin{center}
\begin{tabular}{|c|c|c|c|c|c|c|}\hline
\multicolumn{3}{|c|}{Fig.1 } &\multicolumn{2}{|c|}{Fig. 2} &
\multicolumn{2}{|c|}{Fig. 3 ($\pi^-$)}\\ \hline
$\beta$(\%) & $\eta=2.6$ & $ \eta=3.2$ &
$\beta$(\%)  &  PHENIX & $\beta$(\%) & $y=3$\\ \hline
0-10 & 0.4340 & 0.3448 & 0-20 & 0.563 & 0-20 & 0.3228\\ \hline
10-20 & 0.4359 & 0.3452 & 20-40 & 0.572 & 30-50& 0.3151 \\ \hline
20-40 & 0.4363 & 0.3458 & 40-60 & 0.601 & 60-80 & 0.3108 \\ \hline
40-60 & 0.4366 & 0.3462 & 60-88 & 0.601 & &\\ \hline
\end{tabular}
\end{center}
\caption{Fitted parameter $\langle p_T\rangle$ in the scaling distribution
for Au+Au and d+Au collisions at $\sqrt{s_{NN}}=200$ GeV, for different
centralities and rapidities shown in Figs. (1-3).}
\label{tab1}
\end{table}

\begin{figure}[tbph]
\includegraphics[width=0.4\textwidth]{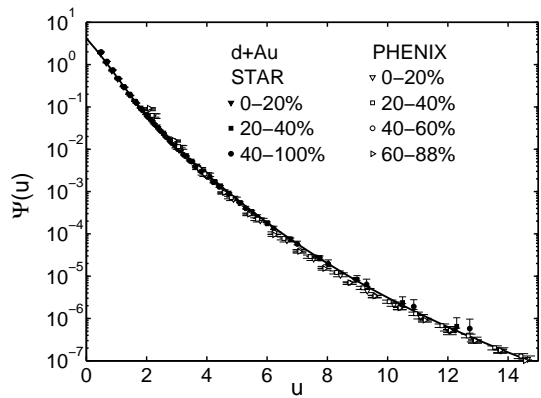}
\caption{Scaling distribution $\Psi(u)$ showing the coalescence spectra
from the STAR data for charged particle and from PHENIX $\pi^0$ data  production in
d+Au collisions at mid-rapidity. The data points are taken from \cite{exp2,exp3}.}
\label{fig2}
\end{figure}

The second observation in this paper is that the same scaling $p_T$ distribution can
describe the $p_T$ distributions of charged particles produced in d+Au collisions
 \cite{exp2} and $\pi^0$ data \cite{exp3}in central rapidity region at different centralities,
as shown in Fig. \ref{fig2}. The fitted values of parameter $\langle p_T\rangle$ for
the PHENIX data are also tabulated in TABLE I.

\begin{figure}[tbph]
\includegraphics[width=0.4\textwidth]{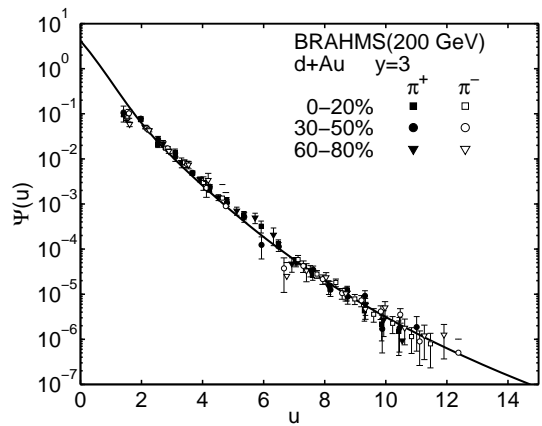}
\caption{Scaling distribution $\Psi(u)$ showing the coalescence of four centrality
bins in the BRAHMS data for $\pi^-$ production in
d+Au collisions at forward direction. The data points are taken from \cite{hongy}.}
\label{fig3}
\end{figure}

To investigate the rapidity dependence of the distribution in d+Au collisions, we
take the data for $\pi^+$ and $\pi^-$ at $y=3$ for different centrality cuts \cite{hongy}
and try to put the data points to the obtained scaling curve with exactly the same
procedure as mentioned above. The agreement is quite good, as shown
in Fig. \ref{fig3}. The fitted values of the parameter $\langle p_T\rangle$
are also given in TABLE I. Then we can conclude that the same scaling function
can describe pion and charged particle production in both Au+Au and d+Au collisions
at different centralities and rapidities.
In this sense, we can say that the obtained scaling distribution is universal.

To show the degree of agreement between the experimental data and the fitted results, we define
$B$ as the ratio of the data to the fitted results. For illustration, $B$, calculated
from results in Fig. 3, as a function of $u$ is shown in Fig.4.
Obviously, values of $B$ are very close to 1, indicating that the scaling of the produced
pion spectrum is true with high accuracy.
\begin{figure}[tbph]
\includegraphics[width=0.4\textwidth]{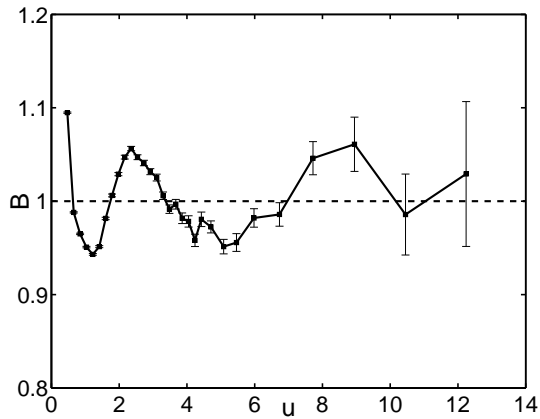}
\caption{Ratio $B$ of the experimental data and the fitted results, as shown in Fig. 3,
for the pion production as forward direction.}
\label{fig4}
\end{figure}

Once we have found a scaling law for the distributions for Au+Au and d+Au
collisions for different centralities and rapidities, we need to dig out the physical
implications behind the scaling behavior. From the scaling function $\Psi(u)$, one
can see that the only parameter characterizing the normalized distribution is
the average transverse momentum $\langle p_T\rangle$ which depends on the centrality,
rapidity and the colliding energy and system. Once $\langle p_T\rangle$ is known, both the
soft part with low $p_T$ and the hard part with high $p_T$ are determined from
$\Psi(u)$. Traditionally it was assumed that the bulk of particles is produced
by soft processes which determine $\langle p_T\rangle$. If the soft
and hard processes can be separated, as used in the two-component model \cite{xnwang},
and each part has different dynamical origin, it is extremely hard to imagine that they
can be united into a single scaling function. Thus, we conclude that the hard process must
contribute considerably to the average $p_T$ and is a coherent part of particle production.
In fact, we have known from the study of particle correlations in central Au+Au
collisions at RHIC energies that the lost energy from hard partons must heat the
medium and increase the mean transverse momentum.

\begin{figure}[tbph]
\includegraphics[width=0.4\textwidth]{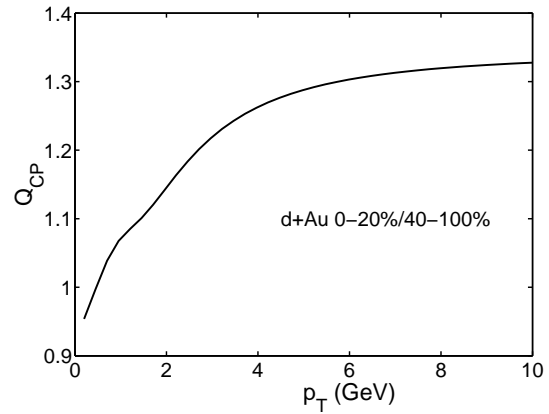}
\caption{$Q_{CP}$ as a function of $p_T$ for d+Au collisions.
The centrality cuts are 0-20\% and 40-100\%, respectively.}
\label{fig5}
\end{figure}

On the other hand, we have known from experiments that the particle production is
suppressed, called jet quenching \cite{quench}, in central Au+Au collisions because
of the possible formation of dense hot medium but enhanced in d+Au collisions due to the
Cronin effect \cite{cronin}, possibly because of the $p_T$ broadening from multiple
soft collisions before a hard one \cite{multi}, or from the recombination of soft
partons with those from showers \cite{recom}. One can show that the enhanced production
in d+Au collisions can be accounted for from our fitted results that the mean transverse
momentum $\langle p_T\rangle$ increases from peripheral to central collisions.
 We can define a new variable $Q_{\rm CP}$
to measure the suppression or enhancement from the scaling function $\Phi(p_T,y,\beta)$ as
\begin{equation}
Q_{\rm CP}=\frac{\Phi(p_T,y,\beta)_C}{\Phi(p_T,y,\beta)_P}=\frac{\left(\Psi(p_T/\langle p_T
\rangle)/\langle p_T\rangle^2\right)_C}{\left(\Psi(p_T/\langle p_T\rangle)
/\langle p_T\rangle^2\right)_P} .
\label{eq2}
\end{equation}
An advantage of this definition of $Q_{\rm CP}$ over the traditional $R_{CP}$ is that it
involves only the probability for a produced particle having certain transverse
momentum and is independent of the total multiplicity. So the overall enhancement
in the backward region ($y<0$) in d+Au collisions cannot be shown in $Q_{\rm CP}$.
In calculating the value of $Q_{\rm CP}$ we do not need to use the number of binary
hard collisions, which is model dependent. $Q_{\rm CP}$ is shown in Fig. \ref{fig5} as a
function of $p_T$ for d+Au collisions at mid-rapidity and it increases slowly to about 1.33.
Thus the production of high $p_T$ pions is enhancement by about 30 percent.
This is in agreement with experimental observations \cite{enhan}.
In this way the enhancement or suppression behavior
can be shown clearly in a model independent way.

Finally, we investigate nucleon-nucleon collisions at $\sqrt{s}=200$ GeV to
see whether the particle spectrum can be described by the same scaling function.
The experimental data on the $p_T$ distribution in the central rapidity region
in p+p collisions can be parameterized \cite{hage} as
\begin{equation}
\frac{dN}{p_Tdp_T}=A(a+p_T)^{-n}\ ,
\end{equation}
where $A, a$ and $n$ can be determined from fitting the data. It is astonishing
that this parameterization for pion distribution in p+p collisions,
with the parameters suitably chosen, can also be put to the scaling curve for
$\Psi(u)$ by shrinking horizontally and shifting vertically. As shown in Fig. 6 the
$\pi^+$ spectrum in p+p collisions can be fitted to the scaling function.
Then we come to our final conclusion that the same scaling
function can be used, for different colliding systems at different centralities and
rapidities, to give the probability for a produced particle having $p_T$
once the parameter $\langle p_T\rangle$ for the average transverse momentum is
known, thus the scaling function is universal.

Since $\Psi(u)$ is universal, the centrality and rapidity dependence
in $\Phi(p_T,y,\beta)$ is totally encoded in $\langle p_T\rangle$.
Straightforwardly, we can have
\begin{eqnarray}
\langle p_T^n\rangle& \equiv &\int p_T^n \Phi(p_T,y,\beta)p_Tdp_T\nonumber\\
& = & \langle p_T\rangle^n \int_0^\infty u^n\Psi(u) udu\ .
\end{eqnarray}
Therefore, the ratio $\langle p_T^n\rangle/\langle p_T\rangle^n$ is a constant
independent of the colliding system, the centrality, and rapidity. This consequence
can be easily checked experimentally. For $n=2, 3$ and 4, the ratio equals to 1.65, 4.08
and 14.39, respectively.

\begin{figure}[tbph]
\includegraphics[width=0.4\textwidth]{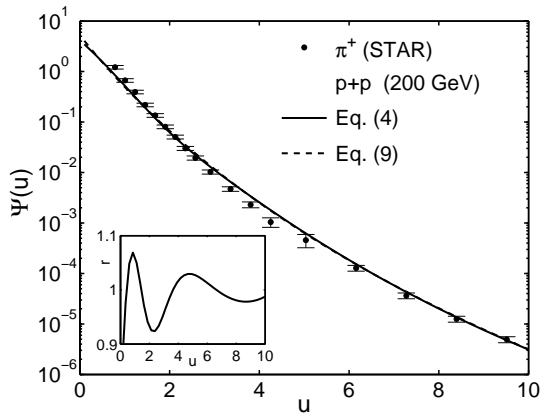}
\caption{Comparison between $\Psi(u)$ from fitting the data and from the maximum entropy
method. The insert gives the ratio between them. The points are
$\pi^+$ distribution in p+p collisions taken from \cite{pipp}.}
\label{fig6}
\end{figure}

We are now in a position to derive an expression for the particle distribution
from some statistical consideration. One method is from the principle
of maximum entropy. The scaling function is not an exponential form, therefore the
well-known Shannon entropy cannot be used for our purpose. Instead, as
suggested in \cite{biro}, a non-extensive entropy may be used, considering the complex
nonlinear interactions and the possible presence of fractal chaotic behavior in nuclear
collisions, to derive an expression for the particle distribution. The entropy we will
use here is the Tsallis entropy \cite{tsi} which reads for our case
\begin{equation}
S_q=(1-\int \Phi^q(p_T) p_Tdp_T)/(q-1)\ ,
\end{equation}
with parameter $q$ as a measure of the non-extensiveness of the system.
The maximum entropy under conditions demanded by Eqs. (\ref{norm}) and (\ref{ave})
gives, after the same shrinking and shifting as above,
\begin{equation}
\Psi(u)=A(a+u)^{-n}\ ,
\label{tsa}
\end{equation}
with $n=1/(1-q)$ and $a=(n-3)/2, A=a^{n-2}(n-1)(n-2)$.
So there is one parameter $q$ to determine the shape of the distribution.
The functional form of $\Psi(u)$ derived from the maximum entropy method
is very different from that given by Eq. (\ref{scal}) but is the same as that used
in parameterizing the p+p data.
Then one can compare the distributions $\Psi(u)$ from our fitting procedure and from
the maximum entropy. As shown in Fig. \ref{fig6}, they agree with each other extremely
well in a very wide range of $u$ when $n=13.2$. There in log scale one can hardly see any
difference between them. Small difference can be seen if we show a ratio between them in a
linear scale. As can be seen from the insert in Fig. 6 the largest difference is less than
ten percent, though the distribution covers six orders of magnitudes in the shown range of
$u$. This agreement indicates that the produced particle system, including both the soft
and hard particles, may be a non-extensive fractal system and the fractal property is
independent the colliding system. The parameter characterizing the fractal property of
the system is then $q=1-1/n=0.924$, not too far from 1.

In summary, a scaling distribution for pions is found for Au+Au, d+Au and p+p collisions
at RHIC energy. The scaling distribution is universal, since it does not depend
on the colliding system, the centrality and rapidity. What characterizes the dynamical
difference in the particle production is the value of the average transverse momentum
$\langle p_T\rangle$. A simple prediction which can be checked experimentally is that
$\langle p_T^n\rangle/\langle p_T\rangle^n$ is independent of the colliding system,
the colliding centrality and rapidity. Possible statistical origin for the scaling
distribution of pions is discussed from a non-extensive entropy. Different particles
may have different scaling properties.

This work was supported in part by the National Natural Science
Foundation of China under grant No. 10475032.

\end{document}